\documentclass[12pt,a4j]{article}

\usepackage{amsmath}
\usepackage{amssymb}
\usepackage{amsthm}
\usepackage{mathrsfs}
\usepackage{color}
\usepackage{fullpage}
\usepackage{docmute}
\usepackage{mathtools}

\usepackage{ulem} 

\usepackage{bm}

\usepackage{here}

\usepackage{indentfirst}

\usepackage{url}

\theoremstyle{break}
\newtheorem{thm}{Theorem}

\newtheorem{prop}{Proposition}
\newtheorem{lem}{Lemma}

\theoremstyle{remark}

\newtheorem{ex}{Example}

\newcounter{numA} 
\newcounter{numB}
\newcommand{\ZZ}{\mathbb{Z}} 
\newcommand{\RR}{\mathbb{R}} 
\newcommand{\CC}{\mathbb{C}} 
\newcommand{\TT}{\mathbb{T}} 

\newcommand{\ol}[1]{\overline{#1}} 
\newcommand{\wt}[1]{\widetilde{#1}} 
\newcommand{\wh}[1]{\widehat{#1}} 
\newcommand{\id}{\mathrm{id}} 

\newcommand{\diam}{\mathrm{diam}} 
\newcommand{\area}{\mathrm{area}} 

\newcommand{\ca}[1]{\mathcal{#1}} 

\def\inprod<#1>{\left\langle #1 \right\rangle} 

\mathtoolsset{showonlyrefs=true}

\title
{ \Large \bf 
 Cover time of graphs with bounded genus
}
\author{Naoki Matsumoto\thanks{Research Institute for Digital Media and Content, 
Keio University, Kanagawa, Japan,
e-mail: \tt{naokimatsumoto@dmc.keio.ac.jp}}
and 
Yuuki Takai\thanks{Academic Foundations Programs, Kanazawa Institute of Technologym,
Kanazawa Institute of Technology, Ishikawa, Japan,
e-mail: \tt{takai@neptune.kanazawa-it.ac.jp}}}
\date{}

\begin{document}

\maketitle

\abstract{
The cover time of a finite connected graph is the expected number of steps needed 
for a simple random walk on the graph to visit all vertices of the graph. 
It is known that the cover time of any finite connected $n$-vertex graph
is at least $(1 + o(1)) n \log n$ and at most $(1 + o(1)) \frac{4}{27} n^3$.
By Jonasson and Schramm, the cover time of any bounded-degree finite connected $n$-vertex planar graph
is at least $c n(\log n)^2$ and at most $6n^2$, 
where $c$ is a positive constant depending only on the maximal degree of the graph. 
In particular, 
the lower bound is established via the use of circle packing of planar graphs on the Riemann sphere.
In this paper, 
we show that the cover time of any finite $n$-vertex graph $G$ with maximum degree $\Delta$
on the compact Riemann surface $S$ of given genus $g$
is at least $c n(\log n)^2/ \Delta(g + 1)$ and at most $(6 + o(1))n^2$, 
where $c$ is an absolute constant,
if $n$ is sufficiently large and three sufficient conditions for $S$ and a circle packing of $G$ filling $S$.
}

\section{Introduction}

A {\it random walk} on a graph is a simple stochastic process such that
whenever a random walker at a certain vertex chooses a neighbor uniformly at random and moves to this neighbor.
The study of random walks ranges over many research fields, e.g., probability, graph theory, and algebra
(for several major topics and well-known results, see books~\cite{aldous2002reversible,woess2000random}).
Moreover,
there are many not only mathematical researches but also applications of random walks so far; 
for example, PageRank~\cite{brin1998anatomy} is the most typical application of random walk based algorithms
(see surveys~\cite{masuda2017random,xia2019random} 
and in particular, the former addresses both discrete and continuous-time random walks).
The main subject in this paper is the {\it cover time} of a finite graph by random walks,
which is the expected number of steps needed 
for a simple random walk on the graph to visit all vertices of the graph~\cite{aldous1989introduction}. 
At the same time, as being an interesting basic concept of random walks,
the cover time of graphs is an important invariant 
for network analysis such as community detection,
since it gives us an amount of time to cover all vertices of a community
(see surveys about community detection by random walks~\cite{bedi2016community,chakraborty2017metrics}).

In what follows, all graphs are finite, simple, and undirected unless we particularly mention them.
Feige estimated the lower and upper bounds of cover times of connected graphs as follows.

\begin{thm}[\cite{feige1995tight_lower,feige1995tight_upper}]\label{thm:bounds}
Let $G$ be a connected graph with $n$ vertices.
Then the cover time of $G$ is at least $(1 + o(1)) n \log n$ and at most $(1 + o(1)) \frac{4}{27} n^3$.
\end{thm}

The bounds of Theorem~\ref{thm:bounds} are the best possible,
and a complete graph and a lollipop graph attain the lower and upper bound, respectively,
where a {\it lollipop graph} is obtained from a complete graph $K$ and a path $P$
by identifying a vertex of $K$ and an end-vertex of $P$.
Since a lollipop graph contains a clique,
it is worth considering the cover time of a graph with a bounded genus of compact Riemann surfaces
(i.e., compact orientable topological surfaces)
on which the graph can be embedded;
it is well known that 
the order of complete graphs which can be embedded on a compact Riemann surface of genus $g$
is bounded from above (by a function depending only on $g$) when $g$ is fixed.
So, we focus on the cover time of graphs $G$ with genus $g$ bounded.

Jonasson and Schramm estimated the cover time of planar graphs 
(i.e., graphs on the Riemann sphere, which is a Riemann surface of genus zero) 
with a bounded maximum degree, as follows.

\begin{thm}[\cite{jonasson2000cover}]\label{thm:jonasson2000}
Let $G$ be a connected planar graph with $n$ vertices and maximum degree $\Delta$.
Then the cover time of $G$ is greater than $cn(\log n)^2$ and less than $6n^2$,
where $c$ is a positive constant depending only on $\Delta$.
\end{thm}

In this paper,
we extend Theorem~\ref{thm:jonasson2000} to a general compact Riemann surface,
as described in the following main theorem.

\medskip
\noindent
{\bf Main theorem.}
Let $\{ G_k =(V_k, E_k) \}_{k\geq 0}$ be an increasing sequence of 
connected graphs with maximum degree $\Delta$ and minimum genus $g$. 
We assume that $\{ G_k \}$ satisfies three assumptions 
(described at the beginning of Section~\ref{sec:proof}).
Then, there is an absolute constant $c$ 
such that for any sufficiently large $k$, the following 
holds:
\[
 \frac{c |V_k| (\log |V_k|)^2}{\Delta(g+1)} < \mbox{cover time of $G_k$} 
 \leq \left( 6+\frac{12g-18}{|V_k|} - \frac{12g-12}{{|V_k|}^2} \right){|V_k|}^2. 
\]
\smallskip

Although there is a nice affinity between random walks and circle packing of planar graphs~\cite{nachmias2020planar}, 
there is no result about random walks using circle packing of graphs on non-spherical Riemann surfaces.
In the proof of Theorem~\ref{thm:jonasson2000},
Jonasson and Schramm applied such a nice relation for planar graphs to estimate the cover time of those graphs.
The lower bound of our main theorem is established using circle packing of graphs on Riemann surfaces,
which is a generalization of Jonasson and Schramm's proof for their lower bound.
For generalization of Jonasson and Schramm's method,
our proof contains several known results established in distinct research fields,
e.g., topology and theory of discrete analytic functions.

The paper is organized as follows.
In the next section, we introduce the precise definitions of notions and notations used in this paper.
In Section~3, 
we give the precise statement of the main theorem and prove it.
In Section~4, 
we remark on the cover time of some graphs of minimum genus $g$.

\section{Preliminaries}

In this section, 
we introduce bare minimum terms and lemmas concerning cover time and circle packing on surfaces
to prove the main theorem.

\subsection{Cover time}

Let $G = (V,E)$ be an undirected finite graph and $\{ X_t \}_{t \geq 0}$ be 
a simple random walk on $G$ such that $X_0 = v$ where $v \in V$.  
We define the {\it first passage time} $T_u^v$ of $\{ X_t \}$ from 
$v$ to $u$ is defined by 
\[
 T_u^v := \min\{ t \geq 0 \mid X_t = u \}. 
\]
The {\it cover time} $C^v$ of the random walk $\{ X_t \}$ is defined by 
\[
 C^v = \max\{ T_u \mid u \in V \}.  
\]
The main object in this paper is the cover time $E_v(C^v)=E_v(C)$, which is 
defined by the expected number of steps that it takes for a random walk that starts 
at $v$ to visit all vertices of the graph. 
We define the cover time $E_G(C)$ of $G$ by $E_G(C)=\max_{v}E_v(C)$. 

To analyze $E_v(C^v)$, we use the following: 
\begin{align}
H(u,v) &= E_v T_u^v : \text{hitting time from } v \text{ to } u, \\ 
C(u,v) &= H(u,v) + H(v,u) : \text{commute time}, \\ 
D(u,v) & = H(u,v) - H(v,u) : \text{difference time}. 
\end{align}
It is known that the triangle equation for $D(u,v)$ holds (\cite[Lemma~2]{coppersmith1993collisions}): 
\begin{align}
 D(u,v) + D(v,w) = D(u,w). \label{eq:triangle-equation-of-D} 
\end{align}
The effective resistance $R(u,v)$ is also important. This  
is defined as 
\[
 R(u,v) = \sup_{f \in \RR^V, \ca{D}(f)\neq 0} \frac{(f(u) - f(v))^2}{\ca{D}(f)}, 
\]
where $\ca{D}(f) = \sum_{ \{u,v\}\in E} (f(u)-f(v))^2$. 
Note that $R(u,v) = R(v,u)$ for any $u,v \in V$.
By \cite[Theorem~2.1]{chandra1996electrical} or \cite[Corollary~2.1]{tetali1991random}, 
the commute time and the effective resistance 
are related by the equation
\begin{align}\label{eq:equation-C-R}
 C(u,v) = 2|E|R(u,v) 
\end{align}
for two vertices $u,v\in V$ on a connected component.  
By using this relation and the triangle inequality of commute times,   
\begin{align}\label{eq:triangle-inequality-R}
 R(u,v) \leq R(u,w) + R(w,v)	
\end{align}
holds for any $u,v,w \in V$. 

By \cite[Theorem~5]{tetali1991random}, the following relation between hitting times and effective resistance is known: 
\begin{align}\label{eq:equation-H-R}
H(u,v) = \frac{1}{2} \sum_{w \in V} d_w (R(u,v) + R(v,w) - R(u,w)),  	
\end{align}
where $d_w$ is the degree of $w$.

As a relation between the expected time of cover times and hitting times, Matthew's inequality is well-known: 
\begin{lem}[{\cite[Theorem~2.6]{aldous2002reversible}}]\label{lem:matthew's-inequality}
Let $G = (V,E)$ be an undirected finite graph and $n = |V|$. 
Then, for any subset $V_0 \subset V$ which is $|V_0|\geq 2$, the following holds: 
\begin{align*}
  h_{|V_0|-1} \min\{ H(u,v) \mid u,v \in V_0, u\neq v \}
  \leq E_G(C) \leq h_{n-1} \max\{ H(u,v) \mid u,v \in V \},
\end{align*}
where $h_m = \sum_{i=1}^m i^{-1}$. 
\end{lem}

\subsection{Circle packing on Surface}


Let $\ca{G}$ be an oriented surface with a metric. 
Then, a set $P = \{c_v \}$ of circles $c_v$ on $\ca{G}$ is said to be 
a {\it circle packing} for a simplicial $2$-complex $K$ if 
\begin{itemize}
 \item[(1)] $P$ has a circle $c_v$ associated with each vertex $v$ of $K$, 
 \item[(2)] two $c_u, c_v$ are externally tangent whenever 
 $\{ u,v \}$ is an edge of $K$, 
 \item[(3)] three circles $c_u, c_v, c_w$ form a positively oriented 
 triple in $\ca{G}$ whenever $\{u,v,w \}$ form a positively 
 oriented face of $K$. 
\end{itemize}

It is known that there is a circle packing for any triangulation of 
the compact orientable topological surface on a Riemann surface as follows: 
\begin{lem}[{\cite[Theorem~4.3]{stephenson2005introduction}}]\label{lem:circle-packing-riemann-surface} 
Let $K$ be a complex that triangulates a compact orientable 
topological surface $S$. 
Then, there exists a Riemann surface $\ca{S}_K$ homeomorphic to $S$ 
and a circle packing $P$ for $K$ in the associated intrinsic spherical, euclidean, or hyperbolic metric on $\ca{S}_K$ such that 
$P$ is univalent and fills $\ca{S}_K$. The Riemann surface $\ca{S}_K$ 
is unique up to conformal equivalence, and $P$ is unique up to 
conformal automorphisms of $\ca{S}_K$.  
\end{lem}


\subsection{Branched covering}

Let $S_1, S_2$ be two Riemann surfaces, and $f\colon S_1 \to S_2$ a nontrivial 
analytic map. 
Then, for any $P \in S_1$ and charts
$z=\varphi_i^1(P) \in \varphi_i^1(U_i^1)$ and $f(P)\in U_j^2$, 
by changing the charts if it is necessary, 
$\varphi_j^2 \circ f\circ (\varphi_i^1)^{-1} (z)$ can be approximated 
as 
\[
 \varphi_j^2 \circ f\circ (\varphi_i^1)^{-1} (z) \sim z^n. 
\] 
Then, $n$ is called by the ramification index of $f$ at $P$, denoted 
by $e_P= e_P(f)$. 
If $e_P >1$,
then the map $f$ is called a {\it branced covering}
and $P\in S_1$ is called a {\it branch point} of $f$.

We assume that $f$ is nontrivial. Then, it is known that for any 
$Q \in S_2$, the summation 
\[
 \sum_{P \in f^{-1}(Q)} e_P 
\] 
is a constant independent of $Q$. We call this summation the {\it degree} of $f$. This is denoted by $\deg(f)$.


Let $S$ be a compact Riemann surface of genus $g$. Then, as a consequence of 
the Riemann-Roch Theorem, we have the following: 
\begin{lem}[{\cite[Corollary~16.12]{forster2012lectures}}]\label{lem:bound-of-degree}
 There is a branched covering $f\colon S \to \wh{\CC}$ whose 
 degree is at most $g+1$. 
\end{lem}


On the other hand, let $S_1, S_2$ be compact Riemann surfaces of genus $g_1, g_2$ respectively. 
Then, the following is known: 
\begin{lem}[{Riemann-Hurwitz formula \cite[\S.17.14]{forster2012lectures}}]\label{lem:Riemann-Hurwitz} 
For any analytic map $f\colon S_1 \to S_2$, the following holds: 
\[
 2-2g_1 = \deg(f)(2-2g_2) - \sum_{P\in S_1} (e_P -1). 
\]
\end{lem}

By Lemma~\ref{lem:Riemann-Hurwitz}, an analytic map $f$ appearing in  
Lemma~\ref{lem:bound-of-degree} satisfies the following.
\begin{lem}
\label{lem:bound-of-num-of-branch-points}
The following holds: 
\[
\#\{\text{\rm branch points of } f \} \leq \sum_{P\in S_1} (e_P -1)  =  2\deg(f) - (2-2g) \leq 4g = O(g).
\]
\end{lem}

\section{Statement and proofs}\label{sec:proof}

\subsection{Statement}


Let $\{G_k=(V_k, E_k) \}_{k \geq 0}$ be an increasing sequence of 
undirected finite connected graphs with maximum degree $\Delta$ 
and minimum genus $g$. 
We assume that $G_k$ satisfies the following assumptions: 
\begin{enumerate}
	\item[(A1)] There is a compact Riemann surface $S_k$ of genus $g$ which is filled by a circle packing $P_k=\{ C_v \mid v\in V_k \}$ of $G_k$. 
	\item[(A2)] The sequence of compact Riemann surfaces $\{ S_k \}$ converges to a compact Riemann surface $S_\infty$ in the moduli of 
	compact Riemann surfaces $X_g$ of genus $g$, and 
	there is an orientation-preserving homeomorphism 
	$\varphi_k\colon S_\infty \to S_k$ 
		which converges to the identity map of $S_\infty$ in the 
		moduli space. 
	\item[(A3)] The maximum of radii of circles in $P_k$ converges to zero when $k$ goes to $\infty$. 
\end{enumerate}

Under these assumptions, the main theorem in this paper is the following: 
\begin{thm}\label{thm:main}
Let $\{ G_k =(V_k, E_k) \}_{k\geq 0}$ be an increasing sequence of 
undirected finite connected graphs with maximum degree $\Delta$ 
and minimum genus $g$. We assume that $\{ G_k \}$ satisfies the assumptions 
{\rm (A1), (A2),} and {\rm (A3)}. 
Then, there is an absolute constant $c$ such that for any sufficiently large $k$, the following 
holds:  
\[
 \frac{c |V_k| (\log |V_k|)^2}{\Delta(g+1)} < E_{G_k}(C) \leq \left( 6+\frac{12g-18}{|V_k|} - \frac{12g-12}{{|V_k|}^2} \right){|V_k|}^2. 
\]
\end{thm}

We show some examples satisfying the assumptions (A1), (A2), and (A3). 
\begin{ex}
 The grid graph $(\ZZ/k\ZZ)^2$ satisfies the assumptions 
 {\rm (A1), (A2),} and {\rm (A3)}. Indeed, we consider the 
 lattice $\Lambda=\ZZ+\ZZ\sqrt{-1}$ in 
 $\CC$ and a grid graph $\Lambda_k =(1/k)\Lambda$. Then, the set of circles 
 $\wt{P}_k = \{ C(z) \mid z \in (1/k)\Lambda \}$, where $C(z)$ is the circle in 
 $\CC$ whose center is $z$ and radius is $1/2k$. Then, the quotient 
 $\TT = \CC/\Lambda$ becomes a torus and $\Lambda_k/\Lambda$ becomes the grid graph 
 $(\ZZ/k\ZZ)^2$. Then, the quotient $P_k$ of the set of circles $\wt{P}_k$ can be regarded as a circle packing of the grid graph on the torus 
 $\TT$. Hence, by replacing $G_k = (\ZZ/k\ZZ)^2$, $S_k=S_\infty=\TT$, 
 $\varphi_k=\id$, the assumption holds for grid graph $(\ZZ/k\ZZ)^2$. 
 We remark that the order of the lower bound of Theorem~\ref{thm:main} 
 is best possible in general
 by the results in \cite{zuckerman1990technique} or \cite{dembo2004cover},
 since a grid graph has maximum degree at most~4. 
\end{ex}

\begin{ex}
 Let $G=G_0$ be a connected finite triangulation with maximum degree $\Delta$ and 
 minimum genus $g$ and $G_k = (V_k, E_k)$ be the graph obtained by taking 
 $k$ times of the hexiagonal refinement of $G$. Then, 
 $\{ G_k \}$ is an increasing sequence of finite graphs with maximum degree 
 $\Delta$ and minimum genus $g$. As in \cite{kelner2006spectral}, we can 
 show that the sequence $\{ G_k \}$ satisfies the assumption (A1), (A2), and 
 (A3). 
\end{ex}

Note that the order of the upper bound of Theorem~\ref{thm:main} is best possible in general
since the cover time of a path with $n$ vertices is $\Omega(n^2)$.

\subsection{The upper bound} 
The upper bound of Theorem~\ref{thm:main} can be proved for any finite graph 
$G$ of minimum genus $g$. 
Hence, we show this for such a graph $G=(V,E)$ with $|V| = n$.

In general, the following estimate of upper bounds has been known: 
\begin{lem}[{\cite[Theorem~1. Chapter~6]{aldous2002reversible}}]\label{lem:aldous}
 Let $\ol{d}$ be the average degree of $G$. Then, the cover 
  time of random walk is bounded from above as follows: 
\[
 E_G(C) \leq \ol{d}n(n-1). 
\]  
\end{lem}

If the graph $G$ is a finite graph of minimum genus $g$, 
then Euler's polyhedron formula implies 
\[
 |V| - |E| + |F| = 2-2g, 
\]
where $F$ is the set of faces of the simplicial complex defined by $G$. Because $|F|\leq 2|E|/3$ holds, we have 
\[
 |V| - \frac{|E|}{3} \geq 2-2g. 
\]
Thus, the average $\ol{d} = 2|E|/|V|$ can be written as 
\[
 \ol{d} \leq 6+ \frac{12(g-1)}{n}.
\]
This implies the following.
\begin{prop}
The following holds: 
\begin{align*}
 E_v(C) &\leq 6n(n-1) + 12(g-1)(n-1) = n^2\left( 6+\frac{12g-18}{n} - \frac{12g-12}{n^2} \right).
\end{align*}
\end{prop}
\qed

\subsection{The lower bound} 


Let $\{ G_k =(V_k, E_k) \}_{k\geq 0}$ be 
an increasing sequence of undirected finite connected graphs with maximum degree $\Delta$ and minimum genus $g$. 
We assume that $\{ G_k \}$ satisfies the assumptions (A1), (A2) and (A3). 
Let $S_k$ be a compact Riemann surface which fills  
a circle packing $P_k=\{ C_v \}_{v \in V_k}$ of $G_k$ by Lemma~\ref{lem:circle-packing-riemann-surface}, 
and $z_v \in S_k$ the center of the circle $C_v$.

By Lemma~\ref{lem:bound-of-degree}, there is a branched covering 
$f_\infty \colon S_\infty \to \wh{\CC}$ 
whose degree is at most $g+1$. We define $f_k = f_\infty\circ\varphi_k^{-1} \colon S_k \to \wh{\CC}$,
where $\wh{\CC} = \CC  \cup  \{\infty\}$.
Then, $f_k$ is homotopic to a quasiconformal map of the same degree 
(denoted by the same symbol $f_k$). We remark that by the assumption (A2), $f_k$ converges to $f_\infty$. 
Then, we show the following inequality: 
\begin{lem}\label{lem:key-inequlity}
For any sufficiently large $k$, 
there is a positive constant $A$ such that for any $w,u \in V_k$ satisfying $\{ w, u\} \not\in E_k$, $f_k(z_w), f_k(z_u)\not\in \{\infty \}$, and $|f_k(z_w)-f_k(z_u)|\geq 3\wt{\delta}$,  
\[
 R_{G_k}(w,u) \geq \frac{A}{\Delta(g+1)} \log(|f_k(z_w) -f_k(z_u)|/r'_u) 
\]
holds. Here, $r'_u$ is the maximum radius of a circle of 
the center $f_k(z_u)$ included in $f_k(C_u)$, 
i.e.,  
\[
 r'_u = \max \{ r >0 \mid B_r(f_k(z_u))\subset f_k(C_u) \}, 
\]
 $\wt{\delta}$ is the maximum of the diameters of 
 $\{F_k(C_v)\}_{v\in V_k}$, i.e., 
\[
 \wt{\delta} = \max\{ \diam(F_k(C_v)) \mid v\in V_k \}, 
\] 
  and  $B_r(z)$ for $z \in \wh{\CC}$ is the closed disk in $\wh{\CC}$ of radius $r$ and of center $z$. 
\end{lem}

\begin{proof}
We fix the vertices $w$ and $u$ as in the statement. 
Then, we define the map 
$F_k\colon S_k{\setminus}f_k^{-1}(\{\infty\}\cup f_k(z_u))\to 
\RR +(\RR/2\pi\ZZ)i$ by $F_k(z)=\log(f_k(z)-f_k(z_u))$. 
Let $\varepsilon$ be a positive constant and 
$K \subset S_k{\setminus}f_k^{-1}(\{\infty\}\cup f_k(z_u))$ a compact subset. 
As an argument in \cite[Section~5]{kelner2006spectral}, 
there are positive numbers $\delta>0$ and $N>0$ such that for any positive number $\delta'<\delta$, 
any $k>N$, any $z\in K$ which is not close to any branch point of $f_k$, 
\[
 H_{F_k}(z;\delta') =  \frac{\max_{d_{S_k}(w, z)=\delta'}|F_k(w)-F_k(z)|}
 {\min_{d_{S_k}(w, z)=\delta'}|F_k(w)-F_k(z)|} \leq 1+\varepsilon
\]
holds, where $d_{S_k}(w,z)$ is the distance between $w$ and $z$ with respect to the metric of constant curvature on $S_k$. 

By taking $k$ as sufficiently large if necessary and the assumption (A3),  
we may assume that any radii of circles in $\{C_v \}_{v\in V_k}$ 
is smaller than $\delta$ by \cite[Lemma~3.4]{bowers2004uniformizing}. 
Under this assumption, for any $v\in V_k \cap K$ such that 
$Disk(C_v) \cap \{ \text{branch points}\}=\emptyset$, 
where $Disk(C_v)$ denotes the disk (on $S_k$) bounded by $C_v$,
the diameter and the area of $F_k(C_v)$ satisfy the following relation: 
\begin{align}\label{eq:diam-area-inequality}
 \diam(F_k(C_v))^2 &\leq (2 \max_{z\in C_v}|F_k(z)-F_k(z_v)|)^2 \notag \\ 
 &\leq 4(1+\varepsilon)^2 (\min_{z\in C_v}|F_k(z)-F_k(z_v)|)^2 \notag \\
 &\leq \frac{4(1+\varepsilon)^2}{\pi}\area(F_k(C_v)).
\end{align}

We set $a:=\log r'_u$, $b = \log |f_k(z_w) - f_k(z_u)|$, and fix a constant
 $c$ such that $c>a+2\wt{\delta}$ and $c<b$, where 
 $\wt{\delta}$ is the maximum of the diameters of 
 $\{F_k(C_v)\}_{v\in V_k}$.
Using the map $F_k$, we define a map 
$g \colon V_k \to \RR$ by 
\[ 
 g(v) := \begin{cases}
 	\min\{ \max\{\text{Re}F_k(z_v), c \}, b \} & \text{if } 
 	v \neq u \\
 	a & 
 	\text{if } v = u. 
 \end{cases}
\]
We use this map $g\colon V_k \to \RR$ to deduce a lower bound of the 
effective resistance 
\begin{align}
 R_{G_k}(w,u) = \sup_{h\colon V_k\to \RR} \frac{(h(w)-h(u))^2}
 {\sum_{ \{v_1, v_2\} \in E_k} (h(v_1)-h(v_2))^2}. \label{eq:effective-resistance}
\end{align}

We divide some cases to give bounds of the summands of the denominator of the right hand side of \eqref{eq:effective-resistance}.  

\begin{enumerate}
\item[(i)]
If $\log|f_k(z_{v_i})-f_k(z_u)|\geq b$ for $i=1,2$ or 
$c\geq \log|f_k(z_{v_i})-f_k(z_u)|$ for $i=1,2$ holds, then 
$|g(v_1)-g(v_2)|=0$ holds. 

\item[(ii)] Suppose that for $\{ v_1, v_2 \} \in E_k$, 
the assumption of (i) does not holds and there is no branch point of $f_k$ 
included in $C_{v_1} \cup C_{v_2}$. 
Then, because there is an element $z' \in C_{v_1}\cap C_{v_2}$, 
 we have 
\begin{align}
 |g(v_1) - g(v_2)| &\leq |g(v_1) - \text{Re}F_k(z')| + |\text{Re}F_k(z')-g(v_2)|\\ 
  &\leq \max_{z \in C_{v_1}} \text{Re}F_k(z) - \min_{z \in C_{v_1}} \text{Re}F_k(z) 
  + \max_{z \in C_{v_2}} \text{Re}F_k(z) - \min_{z \in C_{v_2}} \text{Re}F_k(z) \\ 
  & \leq \diam (F_k(C_{v_1})) + \diam (F_k(C_{v_2})). \label{eq:bound-by-diam}
\end{align}

%

We remark that, $F_k(C_{v_1})$ and $F_k(C_{v_2})$ are included 
in the domain $[c-2\wt{\delta}, b+2\wt{\delta}]\times (\RR/2\pi\ZZ)i$, where 
 $\wt{\delta}$ is the maximum of diameters of $\{F_k(C_v)\}_{v\in V_k}$. 




We set the subset $V_k'\subset V_k$ (resp. $E_k' \subset E_k$) consisting of
the vertices $v\in V_k$ (resp. the edges $\{v_1, v_2 \}\in E_k$) 
such that there is no branch point of $f_k$ 
included in $C_v$ (resp. $C_{v_1}\cup C_{v_2}$).
Then, by the inequality~\eqref{eq:bound-by-diam}, we have 
\begin{align*}
 \sum_{ \{v_1, v_2\} \in E_k'} &(g(v_1)-g(v_2))^2 \\ 
 &\leq 
  \sum_{ \{v_1, v_2\} \in E_k' \atop 
  b+2\wt{\delta} >\text{Re}F_k(z_{v_i})>c-2\wt{\delta}} \{\diam(F_k(C_{v_1})) + \diam(F_k(C_{v_2}))\}^2 \\
 &\leq 
  \sum_{ \{v_1, v_2\} \in E_k' \atop 
  b+2\wt{\delta} >\text{Re}F_k(z_{v_i})>c-2\wt{\delta}} 2\{\diam(F_k(C_{v_1}))^2 + \diam(F_k(C_{v_2}))^2\} \\
 &\leq 
  2\Delta \sum_{v\in V_k' \atop 
  b+2\wt{\delta} >\text{Re}F_k(z_v)>c-2\wt{\delta}} \diam(F_k(C_v))^2. 
\end{align*}

Hence, by the inequality~\eqref{eq:diam-area-inequality}
and applying the inequality \eqref{eq:diam-area-inequality} for 
the compact subset $K=\{ z \in S_k \mid b\geq \text{Re}F_k(z)\geq c \}$,
we have 
\begin{align*}
  2\Delta &\sum_{v\in V_k' \atop 
  b+2\wt{\delta} >\text{Re}F_k(z_{v_i})>c-2\wt{\delta}} \diam(F_k(C_v))^2
  \\
 &\leq 
  \frac{8\Delta(1+\varepsilon)^2}{\pi} \sum_{v\in V_k' \atop 
  b+2\wt{\delta} >\text{Re}F_k(z_{v_i})>c-2\wt{\delta}} \area(F_k(C_v))\\
 &\leq 
 16\Delta(1+\varepsilon)^2 (g+1) (b-c+4\wt{\delta}) 
 \\  &
 <
 16\Delta(1+\varepsilon)^2 (g+1) (b-a+4\wt{\delta}).
\end{align*}
Here, the fifth inequality follows from Lemma~\ref{lem:bound-of-degree}. 

\item[(iii)] The remaining case is that ${v_1, v_2}\in E_k$ does not holds 
the assumption (i) and there are branch points of $f_k$ included in 
$C_{v_1}\cup C_{v_2}$. 

We set $\alpha_i = |f_k(z_{v_i}) - f_k(z_u)|$ for $i=1,2$. 
We may assume that $\log\alpha_2 \geq c$ without loss of generality. 
Then, by the assumption (A3) and the fact that $f_k$ converges to a map $f_\infty$, 
any small constant $\epsilon>0$, 
$|\alpha_1 - \alpha_2|<\epsilon$ holds for sufficiently large $k$. 
(More precisely, $\lim_{k \to \infty}\max|\alpha_1 - \alpha_2| \to 0$ and then $\epsilon \to 0$.)
Then, we have 
\begin{align*}
 (g(v_1) - g(v_2))^2 &= 
 (\log|f_k(z_{v_1}) - f_k(z_{u})| - \log|f_k(z_{v_2}) - f_k(z_{u})|)^2\\ 
 &= (\log(\alpha_1) - \log(\alpha_2))^2\\
 &= \left(\epsilon \frac{1}{\alpha_2} + o(\epsilon)\right)^2
 \leq O(1)\left(\epsilon \frac{1}{e^c}\right)^2. 
\end{align*}
Here, the last inequality follows from $\log\alpha_2 \geq c$. 
Because the degree of $f_k$ is at most $g+1$, 
the number of branch points of $f_k$ is at most $4g$ by Lemma~\ref{lem:bound-of-num-of-branch-points}. 
Hence, the contribution of the set $E_k{\setminus}E_k'$ for 
the denominator of \eqref{eq:effective-resistance} is $O(g)\epsilon^2$. 

\end{enumerate}

Since $f_k$ (hence $F_k$) converges to 
a map as in \cite[Lemma~5.4]{kelner2006spectral} and the radii of circle packings goes to zero when 
$k\to \infty$, the constant $\wt{\delta}$ is $o(1)$ when $k\to \infty$. 
Thus, by combining (i), (ii), (iii), and definition of $R_{G_k}(w,u)$, 
the required inequality  
\begin{align*}
 R_{G_k}(w,u) &\geq \frac{(b-a)^2}{16\Delta (1+\varepsilon)^2(g+1)(b-a+4\wt{\delta})+O(g)\epsilon^2}\\ 
 &\geq \frac{(b-a)^2}{16\Delta (1+\varepsilon)^2(g+1)(b-a+o(1))+o(1)}\\ 
 &\geq \frac{C}{16\Delta (1+\varepsilon)^2(g+1)}(b-a)\\ 
 &\geq \frac{C}{16\Delta (1+\varepsilon)^2(g+1)}\log(|f_k(z_w)-f_k(z_u)|/r'_u) 
\end{align*}
holds for a positive constant $C$. 
\end{proof}

In this situation, we can guarantee an existence of a subset $Z$ 
that guarantees large effective resistance.

\begin{prop}\label{prop:lower-bound-of-resistance-by-order}
For sufficiently large $k$, there are positive constants $A'$ (same as in Lemma~$\ref{lem:key-inequlity}$) and 
$c$ such that for any subset $W \subset V_k$, there is a subset $Z \subset W$ 
 satisfying $|Z|\geq |W|^c/(g+1)$ and for any 
 $u,w \in Z$, 
 \[
  R_{G_k}(u,w) \geq \frac{A'}{\Delta(g+1)} c \log |W|. 
 \]
\end{prop}

\begin{proof} 
We set $n=|W|$. 
Because the map $f_k\colon S_k \to \wh{\CC}$ is of degree at most $g+1$, there is an 
open set $U \subset S_k$ such that 
\[
 |\{v\in W \mid z_v \in U \}| \geq \frac{|W|}{g+1}. 
\]
We fix the open set $U$ and set 
\[
 W_U := \{v\in W \mid z_v \in U \}. 
\]


Let $s$ be a positive number. For this $s$, we define  
\[
 W_j := \{ v \in W_U \mid r_v' \in (n^{s(j-1)}, n^{sj}] \} \quad (j=1,2,\dots), 
\]
where $r'_v$ is same as in the statement of Lemma~\ref{lem:key-inequlity}, i.e., it is defined by 
\[
 r'_v = \max \{ r >0 \mid B_r(f_k(z_v))\subset f_k(C_v) \}.
\]
Then, $\bigsqcup_{j\in \ZZ} W_j = W_U$ holds. 
If $u \in W_j$ and $v\in W_k$ for $k-j \geq 2$, then 
\[
 \frac{r'_v}{r'_u} \geq \frac{n^{s(k-1)}}{n^{s^j}} =n^{s(k-j-1)}\geq n^s
\]
holds. 
Thus, by Lemma~\ref{lem:key-inequlity}, we have 
\[
 R_{G_k}(u,v) \geq \frac{A}{\Delta(g+1)}\log(|f_k(z_u)-f_k(z_v)|/r'_u) 
 \geq \frac{A}{\Delta(g+1)} \log(r'_v/r'_u) \geq 
 \frac{As}{\Delta(g+1)}\log n.
\]
Here, the second inequality follows from the facts that $z_u$ and $z_v$ are in the same open set $U$, 
hence $f_k(z_u) \not \in f_k(C_v)$. 


Now, either of $|\bigsqcup_{j\colon \text{odd}} W_j| \geq |W|/(2(g+1))$ or 
$|\bigsqcup_{j\colon \text{even}} W_j| \geq |W|/(2(g+1))$ holds. 
We assume that the latter case holds.

From now on, we extract a subset $Z_j\subset W_j$ such that for any 
$u,v \in Z_j$ satisfying $u\neq v$, $R_{G_k}(u,v) \geq \frac{A}{\Delta(g+1)}s\log n$ holds.  
We take $Z_j$ as a maximal subset of $W_j$ such that for any $u,v\in Z_j$ satisfying $u\neq v$, 
\begin{align}\label{eq:bounds-of-distance-by-nsj}
 |f_k(z_u) - f_k(z_v)| \geq (1+\varepsilon)n^{s(j+1)}. 
\end{align}


Then, if $u,v\in Z_j$ and $u\neq v$, we have 
\begin{align*}
 R_{G_k}(u,v) &\geq \frac{A}{\Delta(g+1)}\log(|f_k(z_u) - f_k(z_v)|/r'_v) \\
 &\geq \frac{A}{\Delta(g+1)}\log((1+\varepsilon)n^{s(j+1)}/n^{sj})
 =\frac{A}{\Delta(g+1)}\left(s\log n + \log(1+\varepsilon) \right).
\end{align*}


We shall show the lower bound of the order of the obtained subset $Z_j$. 
For a given $v\in W_j$, we consider the circle $C'_v$ in $U$
of radius $2(1+\varepsilon)n^{s(j+1)}$ of center $f_k(z_v)$. 
If for another vertex $u\in W_j$, $f_k(C_u)\not\subset C'_v$ holds, 
then $z_u$ satisfies the inequality~\eqref{eq:bounds-of-distance-by-nsj}. 
Indeed, if $f_k(C_u)\not \subset C_v'$ holds, then 
\begin{align*}
  |f_k(z_u)-f_k(z_v)| &\geq 2(1+\varepsilon)n^{s(j+1)}-\max\{|z-f_k(z_u)| \mid z\in f_k(C_u)\} \\
 & \geq  2(1+\varepsilon)n^{s(j+1)}-(1+\varepsilon)n^{sj} \geq (1+\varepsilon)n^{s(j+1)}
\end{align*}
holds. Here, the second inequilty follows from 
\[
  \max\{|z-f_k(z_u)| \mid z\in f_k(C_u)\} \leq r'_v(1+\varepsilon)\leq n^{sj}(1+\varepsilon)
\]
and the third inequality follows from that $n^{sj}\leq n^{s(j+1)}$. 
This circle $C_v'$ includes at most $\pi(2(1+\varepsilon)n^{s(j+1)})^2/\pi n^{2s(j-1)}=4(1+\varepsilon)^2n^{4s}$ circles corresponding to vertices of $W_j$. 
This implies that the order of a maximal subset $Z_j$ satisfies the inequality 
\begin{align}\label{eq:inequality-of-Z-and-W}
 4(1+\varepsilon)^2n^{4s}|Z_j|\geq |W_j|. 
\end{align}
We set $Z = \bigcup_{j\colon \text{even}}Z_j$. Then, 
\begin{align*} |Z| = \sum_{j\colon \text{even}}|Z_j| &\geq 
 \frac{\sum_{j\colon \text{even}}|W_j|}{4(1+\varepsilon)^2n^{4s}}
 \geq \frac{|W|/2(g+1)}{4(1+\varepsilon)^2n^{4s}}\\ 
 &\geq \frac{n^{1-4s}}{8(1+\varepsilon)^2(g+1)} \geq \frac{n^{1-5s}}{g+1}
\end{align*}
holds for sufficiently large $n$. 
Hence, if we set $s=1/6$, then the statement holds for $c=1/6$.
\qedhere


\end{proof}

Using these lemmas, we deduce a lower bound of the cover time of 
simple random walk on $G_k$.  

\begin{proof}[Proof of the lower bound of Theorem~$\ref{thm:main}$] 
Let $a \in V_k$ be a fixed vertex and set $n_k = |V_k|$.  
We order $\{v_1, \dots, v_{n_k} \}$ as if $i\leq j$, 
\[
 D(a, v_i) \leq D(a, v_j). 
\]
By the triangle equation~\eqref{eq:triangle-equation-of-D}, we have
\[
 D(v_i, v_j) = D(v_i, a) + D(a, v_j) = D(a, v_j) - D(a, v_i). 
\] 
Thus, by the definition of ordering of $\{v_i \}$, $D(v_i, v_j) \geq 0$ holds for any pair $(i,j)$ such that $i\leq j$. 
This together with the definition of difference time implies that 
$H(v_i, v_j) \geq H(v_j, v_i)$ holds if $i\leq j$.

We set $l := \lfloor n_k/2 \rfloor$. We divide the argument as follows: 
\begin{enumerate}
	\item[(i)] There is a pair $(i, j)$ such that $i<j$ and 
	$H(v_j, v_i)\geq n_k(\log n_k)^2/ 2(g+1)$
	\item[(ii)] $H(v_j, v_i) < n_k(\log n_k)^2/ 2(g+1)$ for any pair $(i,j)$ such that $i < j$,
	\begin{enumerate}
	\item $D(v_1, v_l) \geq n_k (\log n_k)^3$, 
	\item $D(v_1, v_l) < n_k (\log n_k)^3$. 
	\end{enumerate}
\end{enumerate}

\noindent
Case (i): We assume that there are $i, j$ such that $i<j$ and 
$H(v_j, v_i)\geq n_k(\log n_k)^2/ 2(g+1)$. 
Then, by Lemma~\ref{lem:matthew's-inequality} and the fact $H(v_i,v_j)\geq H(v_j, v_i)$, we have a claimed lower bound as  
\begin{align*}
	E_v(C^v) \geq \min\{ H(v_j, v_i), H(v_i, v_j) \} 
	= H(v_j, v_i) \geq \frac{n_k(\log n_k)^2}{2(g+1)}. 
\end{align*} 

\noindent 
Case (ii)-(a): We assume that $H(v_j, v_i) < n_k(\log n_k)^2/ 2(g+1)$ holds 
for any pair $(i,j)$ such that $i < j$. 
Moreover, we also assume that $D(v_1, v_l) \geq n_k (\log n_k)^3$.  
Then, for $j\geq l$, we have 
\begin{align*}
	D(v_1, v_j) \geq D(v_1, v_j) + D(v_j, v_l) 
	= D(v_1, v_l) \geq n_k(\log n_k)^3. 
\end{align*}
Here, the first inequality follows from $D(v_j, v_l) \leq 0$ 
and the equality follows from the triangle equation~\eqref{eq:triangle-equation-of-D}. 
Then, $H(v_{n_k}, v_1)$ can be estimated as follows: 
\begin{align*}
 H(v_{n_k}, v_1) &= \frac{1}{2} \sum_{w\in V_k} d_w (R(v_1, v_{n_k}) + R(v_1, w) -R(v_{n_k},w))\\ 
 &\geq \frac{1}{2} \sum_{j=l}^{n_k} (R(v_1, v_{n_k}) + R(v_1, v_j) 
 - R(v_{n_k}, v_j)) \\ 
 & = \frac{1}{4|E_k|} \sum_{j = l}^{n_k} (C(v_1, v_{n_k}) + C(v_1, v_j) 
 - C(v_{n_k}, v_j))
\end{align*}
holds. Here, the first equality follows from the equation~\eqref{eq:equation-H-R}, 
the first inequality follows from the positivity of $R(v_1, v_{n_k}) + R(v_1, w) -R(v_{n_k},w)$ 
proved by the triangle inequality~\eqref{eq:triangle-inequality-R}, 
and the second equality follows from the equation~\eqref{eq:equation-C-R}.

Furthermore, we have 
\begin{align}
 \frac{1}{4|E_k|}& \sum_{j = l}^{n_k} (C(v_1, v_{n_k}) + C(v_1, v_j) 
 - C(v_{n_k}, v_j))\\ 
 & \geq  \frac{1}{12(n_k-2+2g)}
 \sum_{j = l}^{n_k} (C(v_1, v_{n_k}) + C(v_1, v_j) - D(v_j, v_{n_k})
 - 2H(v_{n_k}, v_j)) \label{eq:bound-of-commute-time-1st}\\ 
 & > \frac{1}{12(n_k-2+2g)}
 \sum_{j = l}^{n_k} \left(2D(v_1, v_j) - \frac{n_k(\log n_k)^2}{g+1}
 \right) \label{eq:bound-of-commute-time-2nd}\\ 
 &\geq \frac{1}{12(n_k-2+2g)}
 \sum_{j = l}^{n_k} \left(2n_k(\log n_k)^3 - \frac{n_k(\log n_k)^2}{g+1}\right) \\ 
 &= \frac{1}{12(n_k-2+2g)}
  n_k(\log n_k)^2\left(2\log n_k - \frac{1}{g+1}\right)(n_k-l+1) \\
 &\geq \frac{n_k}{24(n_k-2+2g)}
  n_k(\log n_k)^2\left(2\log n_k - \frac{1}{g+1}\right) \label{eq:bound-of-commute-time-4th}\\
 & \geq \frac{1}{\gamma}
 n_k(\log n_k)^3 \label{eq:bound-of-commute-time-last} 
\end{align}
where $\gamma$ is some constant with $\gamma \geq \frac{24(n_k-2+2g)}{n_k}$.
Here, the first inequality \eqref{eq:bound-of-commute-time-1st} follows from $|E_k|=3n_k + 6g-6$ and  
$D(v_j, v_{n_k})-C(v_{n_k},v_j)=-2H(v_{n_k}, v_j)$, 
the second inequality~\eqref{eq:bound-of-commute-time-2nd} follows from the inequality
\begin{align*}
 C(v_1, v_{n_k}) &+ C(v_1, v_j) - D(v_j, v_{n_k}) \\
 &= H(v_1, v_{n_k})+H(v_{n_k}, v_1) +  H(v_1, v_j)+H(v_j, v_1) - 
 (D(v_j, v_1)+D(v_1, v_{n_k}))\\ 
 &= H(v_1, v_{n_k})+H(v_{n_k}, v_1) +  H(v_1, v_j)+H(v_j, v_1) \\ 
 & \hspace{100pt} -(H(v_j, v_1)-H(v_1, v_j) + H(v_1, v_{n_k})-H(v_{n_k},v_1))\\ 
 &= 2H(v_{n_k}, v_1) + 2H(v_1, v_j)\\ 
 &=2H(v_1, v_j) - 2H(v_j,v_1) + 2(H(v_{n_k},v_1) + H(v_j, v_1))\\ 
 &\geq 2H(v_1, v_j) - 2H(v_j,v_1) \\ &= 2D(v_1, v_j)
\end{align*}
and the assumption $H(v_{n_k}, v_j) < n_k(\log n_k)^2/(2g+2)$,
the fourth inequality \eqref{eq:bound-of-commute-time-4th} follows from $n_k - l + 1 \geq n_k/2$, 
and the 
last inequality~\eqref{eq:bound-of-commute-time-last} 
holds for any $n_k$ satisfying $\log n_k \geq 1/(g+1)$. 
Consequently, the estimate 
\[
 H(v_{n_k}, v_1) \geq \frac{1}{\gamma} n_k (\log n_k)^3
\]
for any sufficiently large $n_k$ contradicts the assumption 
$H(v_j, v_i) < n_k(\log n_k)^2/ 2(g+1)$ for any pair $(i,j)$ such that $i < j$. \\ \par 


\noindent 
Case (ii)-(b): We assume that $H(v_j, v_i) < n_k(\log n_k)^2/ 2(g+1)$  
for any pair $(i,j)$ such that $i < j$ and $D(v_1, v_l) < n_k (\log n_k)^3$ hold. 
Let $W =\{v_1, \dots, v_l \}$ and $Z \subset W$ be the subset in 
Proposition~\ref{prop:lower-bound-of-resistance-by-order}. 
We set $m=|Z| \geq l^c/(g+1)$ and $i_1< i_2< \cdots < i_m$ as $v_{i_j} \in Z$. 
Let $s:= \lfloor \sqrt{m} \rfloor - 1$. 
Then, we have 
\begin{align*}
	\sum_{j=1}^{s-1} D(v_{i_{js}}, v_{i_{(j+1)s}}) = 
	D(v_{i_s}, v_{i_{s^2}}) \leq D(v_1, v_l) < n_k (\log n_k)^3. 
\end{align*}
Here, the first equality follows from the triangle equality~\eqref{eq:triangle-equation-of-D}. 

Here, the first inequality is shown as follows: By the ordering 
$D(a, v_1) \leq D(a, v_{i_s}) \leq D(a, v_{i_{s^2}}) \leq D(a, v_l)$, the triangle equation \eqref{eq:triangle-equation-of-D}, and $D(u,v) = -D(v,u)$, we have 
\[
D(v_1, v_l) = D(a, v_l) - D(a, v_1) \geq D(a, v_{i_{s^2}}) - D(a, v_{i_s}) = D(v_{i_{s}}, v_{i_{s^2}}).  
\]
Thus, there is $t \in \{1, 2, \dots, s-1 \}$ such that 
\[
 D(v_{i_{ts}}, v_{i_{(t+1)s}}) < \frac{n_k(\log n_k)^3}{s-1} = o(n_k). 
\]
Here, the last equality follows from $s \sim n_k^{c/2}$ for the positive number $c$. 
However, if we set 
$V' := \{ v_{i_{ts}}, v_{i_{(ts+1)}}, \dots, v_{i_{(t+1)s}} \} \subset W$, then
\[
 C(u,w) = 2|E_k| R(u,w) \geq \frac{A' n_k \log n_k}{2\Delta(g+1)}
\]
holds for $u,w \in V'$ by Proposition~\ref{prop:lower-bound-of-resistance-by-order} and $|E_k|\geq n_k/2$ 
because $G_k$ is connected (where $A'$ is some constant). 
Hence, we have 
\[
 2H(u,w) = C(u,w) - D(w,u) \geq \frac{A' n_k \log n_k}{2\Delta(g+1)} + o(n_k). 
\]
By applying Lemma~\ref{lem:matthew's-inequality} for $V'$, because 
$h_{|V'|-1} = h_{s} \sim \log s = (c/2)\log n_k - (c/2)\log(g+1)$,  
we obtain the claimed estimate. \qedhere

\end{proof}

\section{Remark on the cover time of graphs with minimum genus $g$} 

It is NP-complete to determine whether 
for a given graph $G$ and a natural number $k$, $G$ has genus at most $k$~\cite{thomassen1989graph}.
Moreover, 
it is hard to exactly compute the cover time of a given graph $G$
even if $G$ is a tree~\cite{higuchi2010exact}.
So it is very difficult
to find a graph exactly attaining the lower bound of Theorem~\ref{thm:main}.
However,
it is known that 
both invariants of a random graph become some values 
{\it with high probability} (for short, whp), as follows.
Let $\mathcal{G}(n, 1/2)$ denote the set of graphs with $n$ vertices 
where each edge occurs independently with probability $1/2$.
\begin{itemize}
\item[(i)] 
For $G \in \mathcal{G}(n, 1/2)$,
the minimum genus of $G$ is $n^2 / 24$ whp~\cite{archdeacon1995genus}.
\item[(ii)] For $G \in \mathcal{G}(n, 1/2)$, $E_G(C) \sim n \log n$ whp
(as a corollary of the result in \cite{jonasson1998cover}).
\end{itemize}

Moreover,
it is well-known that 
for $G \in \mathcal{G}(n, 1/2)$, the maximum degree of $G$ is about $\log n$ whp.
By combining the above values and Theorem~\ref{thm:main},
we can calculate a lower bound,
but it is far from the estimation in (ii).
(Similarly, for random geometric graphs with $n$ vertices,
its cover time becomes about $n \log n$ whp~\cite{cooper2011cover}.
However, we suspect that this is also far from the lower bound of Theorem~\ref{thm:main}.)

Although it is hard to exactly compute the cover time of a given graph,
there are infinitely many $n$-vertex graphs $G$ with minimum genus $g$ and small maximum degree
such that its cover time can be bounded by a function $f(n)$ not depending on $g$.
Here we give one example of such graphs below.

Prepare an $(n-4g)$-vertex tree $T$ with maximum degree~3 and the number of leaves is at least $g$.
(Note that such a tree exists by considering a subgraph of complete binary tree of height $\log n$.)
Let $l_1,l_2,\dots,l_g$ be distinct $g$ leaves of $T$
and let $H_1,H_2,\dots,H_g$ be $g$ copies of the complete graph of order $5$.
Then, for each $i \in \{1,2,\dots,g\}$, we identify $l_i$ and a vertex of $H_i$.
The resulting graph $G$ has minimum genus $g$
since the graph $G'$ obtained from $G$ by contracting all edges of $T$
can be also obtained from $H_1,H_2,\dots,H_g$
by selecting one vertex for each $H_i$ and identifying them,
then by a well-known result in~\cite{battle1962additivity}, we have 
\[
\gamma(G') = \sum_{\mbox{$B$\;:\;block of $G'$}} \gamma(B) = \sum^{g}_{j = 1} \gamma(H_j) = g = \gamma(G)
\]
where a {\it block} is a maximal subgraph without a cut vertex
and $\gamma(H)$ is the minimum genus of a graph $H$.
(Note that the minimum genus of the complete graph of order~5 is one.)
Moreover, $G$ has $n$ vertices and maximum degree~5, and hence, by Lemma~\ref{lem:aldous}, we have
\[
E_G(C) \leq \ol{d}n(n-1) < 5n(n-1),
\]
where $\ol{d}$ is the average degree of $G$.


\bibliographystyle{alpha}
\bibliography{main}

\end{document}